\documentclass[aps,preprint]{revtex4} 

\usepackage{epsf}

\begin{document}

\title{Coupled Cluster Method Calculations Of Quantum Magnets With Spins Of 
 General Spin Quantum Number}

\author{{\underline{D. J. J. Farnell$^a$}}, 
 R. F. Bishop$^b$, and K. A. Gernoth$^b$ \\ }

\affiliation{$^a$School of Mechanical Engineering, \\ 
 University of Leeds, Woodhouse Lane, \\
 Leeds LS2 9JT, United Kingdom\footnote{Email: d.j.j.farnell@leeds.ac.uk ~ Tel: +44 (0)113 233 2179 ~ 
  FAX: +44 (0)113 233 2150}\\}

\affiliation{$^b$Department of Physics, \\
  University of Manchester Institute of
  Science and Technology (UMIST), \\
  P. O. Box 88, Manchester M60 1QD, United Kingdom}

\date{\today}

%
%
%
%
%

\begin{abstract}
We present a new high-order coupled cluster 
method (CCM) formalism for the ground states of lattice quantum spin 
systems for general spin quantum number, $s$.
This new ``general-$s$'' formalism is found to be highly suitable for
a computational implementation, and the technical details of this 
implementation are given.
To illustrate our new formalism we perform 
high-order CCM calculations for the one-dimensional 
spin-half and spin-one antiferromagnetic {\it XXZ} models and 
for the one-dimensional spin-half/spin-one ferrimagnetic {\it XXZ} model. 
The results for the ground-state properties of the isotropic points 
of these systems are seen to be in excellent quantitative 
agreement with exact results for the special case of the spin-half 
antiferromagnet and results of density matrix renormalisation
group (DMRG) calculations for the other systems. 
Extrapolated CCM results for the sublattice 
magnetisation of the spin-half antiferromagnet closely
follow the exact Bethe Ansatz solution, which contains an 
infinite-order phase transition at $\Delta=1$. 
By contrast, extrapolated CCM results for the sublattice 
magnetisation of the spin-one antiferromagnet using this same scheme
are seen to go to zero at $\Delta \approx 1.2$, which is in excellent 
agreement with the value for the onset of the Haldane phase for 
this model.
Results for sublattice magnetisations of the ferrimagnet
for both the spin-half and spin-one spins are non-zero and
finite across a wide range of $\Delta$, up to and including
the Heisenberg point at $\Delta=1$. \\

{\bf Running Head:} CCM Calculations of Quantum Magnets with General Spin

\end{abstract}

\maketitle

\section{Introduction} 

In this article we consider lattice quantum spin systems at zero 
temperature. Since the inception of this subject via the introduction 
of the Heisenberg model it has undergone a period of steady 
development. 
Although the exact Bethe Ansatz solution of the Heisenberg 
model for the spin-half linear chain \cite{ba1,ba2,ba3,ba4} 
followed quickly after its initial introduction, interest 
in lattice quantum spin systems has not faltered ever since.
In particular, exact solutions, such as those typified by the Bethe 
Ansatz, are generally restricted to unfrustrated systems of low-dimensional 
lattices and to low spin quantum number. Thus, much effort has 
been expended over the last fifty or so years in trying to 
understand the properties of more general lattice quantum 
spin systems at zero temperature via approximate methods. 

A recent and exciting topic in this field has been the 
development of the density matrix renormalisation group 
(DMRG) \cite{DMRG1,DMRG2,DMRG3,DMRG4,DMRG5} method. This 
technique allows one to perform highly accurate calculations 
for both frustrated {\it and} unfrustrated systems of general 
spin quantum number. As long as one is able to partition a one-dimensional 
quantum system into ``system'' and ``environment'' subsystems 
then one ought in principle to be able to apply this powerful method of 
modern-day quantum theory. In particular, DMRG calculations have 
been used to demonstrate conclusively that the 
spin-one Heisenberg linear chain antiferromagnet 
contains an excitation gap \cite{Haldane}.
This is in stark contrast to both its classical behaviour and the
behaviour of its quantum spin-half counterpart. However, a major 
restriction on the DMRG method is that it has, so far, only been
conclusively applied to one-dimensional or quasi-one-dimensional
systems, although isolated cases of highly successful DMRG
calculations for various two-dimensional lattices have 
been performed. 

By contrast, quantum Monte Carlo (QMC) 
calculations \cite{qmctheory1,qmctheory2,qmctheory3,qmctheory4} have 
been already applied with much success to spin systems of spatial 
dimensionality both equal to and greater than one. In particular, QMC 
calculations for the zero-temperature properties of the spin-half, 
square-lattice \cite{qmc1,qmc2,qmc3,qmc4} Heisenberg model present 
a very accurate and valuable benchmark against which other 
approximate methods may test themselves. However, although 
QMC is not as limited by spatial dimensionality as the DMRG 
method, it is limited in its range of applicability by the
existence of the well-known ``minus-sign problem.'' For lattice quantum spin 
systems the minus sign problem is, in turn, often a consequence of frustration. 
We note however that for non-frustrated systems one 
can often, but not always, determine a ``sign rule'' \cite{sign_rules1,sign_rules2}
which effectively circumvents the minus-sign problem of QMC. 

Other approximate methods that have been applied to 
lattice quantum spin problems (at zero temperature) include spin-wave 
theory (SWT) \cite{swt1,swt2,SWT}, exact diagonalisations 
of finite-sized lattices \cite{finite1,finite2}, cumulant series 
expansions \cite{series1,series2,series3}, and the coupled 
cluster method (CCM) 
\cite{newccm1,newccm2,newccm3,newccm4,newccm5,newccm6,newccm7,newccm8,newccm9,ccm1,ccm2,ccm999,ccm3,ccm4,ccm5,ccm6,ccm7,ccm8,ccm9,ccm10,ccm11,ccm12,ccm13,ccm14,ccm15,ccm16,ccm17,ccm,ccm18,ccm19}. 
For SWT one maps the spin system onto an exact (or effective) 
bosonic system (for example, via the Holstein-Primakov transformation)
which may then be solved. The advantage of {\it linear} SWT is that 
it is simple to apply and an exact solution generally exists
via the Bogoliubov transformation. Higher-order SWT becomes 
rapidly more complicated and may only be solved perturbatively 
for these higher orders. Also, it is difficult to increase 
the level of approximation of this method in a structured 
and well-defined manner. By contrast, for the QMC method
one may systematically increase the lattice size and still 
obtain an ``exact'' result to within a well-defined statistical 
error. Exact diagonalisations of finite-sized systems 
\cite{finite1,finite2} have also been applied to quantum spin 
models with much success. However, such calculations rapidly become 
limited as the size of the lattice is increased by the amount of 
the computational resources available. By contrast, this problem
is not so evident for the exact cumulant series expansion 
\cite{series1,series2,series3} calculations. Indeed, such calculations 
have provided very accurate results for the ground- and excited-state 
properties of various spin systems. However, in order to determine 
such expectation values a resummation of the otherwise 
rapidly divergent perturbation series must be performed
using Pad\'e approximants or related techniques. Finally, 
a technique of quantum many-body theory (QMBT) called the coupled 
cluster method (CCM) \cite{newccm1,newccm2,newccm3,newccm4,newccm5,newccm6,newccm7,newccm8,newccm9}
has also been applied \cite{ccm1,ccm2,ccm999,ccm3,ccm4,ccm5,ccm6,ccm7,ccm8,ccm9,ccm10,ccm11,ccm12,ccm13,ccm14,ccm15,ccm16,ccm17,ccm,ccm18,ccm19} 
to both unfrustrated and frustrated lattice quantum spin systems 
with considerable success, and it is this method upon which we shall 
concentrate henceforth. 

The first application \cite{ccm1} of the CCM to the  
lattice quantum spin systems was to the spin-half Heisenberg 
model and related models on various bipartite lattices 
using several localised approximation schemes. This work was quickly 
generalised \cite{ccm2,ccm999} to include anisotropy within 
the Hamiltonian via the spin-half {\it XXZ} model. In 
particular, critical points as a function of the anisotropy 
were discovered in an approximation scheme denoted as SUB2 
(in which {\it all} possible two-spin correlations are 
retained). It was furthermore found at 
these critical points that the CCM excitation spectra became ``soft'' 
and that the behaviour with respect to spatial separation
of the spin-spin correlation functions changes from exponential
to algebraic decay. Both of these features were clearly 
strong indicators that these critical points were reflections 
of quantum phase transitions in the real system, which in turn were 
driven by the anisotropy parameter within the  
Hamiltonian. (An interesting application of the {\it extended}
coupled cluster method (ECCM) to this model was also 
performed, and the interested reader is referred to
Ref. \cite{ccm15}.) Frustrated quantum spin systems, 
such as the $J_1$--$J_2$ model \cite{ccm4,ccm5,ccm6,ccm13} 
and the triangular lattice antiferromagnet \cite{ccm11,ccm12}, 
have also been studied using the CCM with equal success. 
Indeed, the CCM has been utilised not only to provide 
accurate values of the ground-state properties of these 
systems and very accurate predictions of their phase 
diagrams, but also to simulate the nodal surfaces 
\cite{ccm13,ccm14} of these models. We note that such 
approximately determined nodal surfaces are potentially 
of very great use to QMC calculations in which the ``sign problem'' is 
present. Other such applications of note of the CCM have 
been to the biquadratic model \cite{ccm3} and to the 
one-dimensional spin-half $J_1$--$J_2$ model using a 
``dimerised'' model state \cite{ccm5}. 

Furthermore, recent high-order CCM calculations 
\cite{ccm7,ccm8,ccm8,ccm9,ccm10,ccm12,ccm13,ccm16,ccm17,ccm,ccm18,ccm19}
for spin-half systems using localised approximation 
schemes have been shown to be very effective in simulating
the ground- and excited-state properties of quantum
spin systems. The strength of this approach is that one
can systematically and rigorously increase the approximation
level as a function of some parameter which increases
the size of the ``locale'' over which one retains {\it all}
multi-spin correlations. The approximation schemes
become exact in the limit of this parameter going to 
infinity, although this is generally impossible to
achieve in a practical application. One can however
extrapolate \cite{ccm16} a series of such results to this limit,
and thus obtain very accurate approximate results  
for the exact properties of even highly frustrated systems.
Indeed, the results produced for both unfrustrated systems 
such as the spin-half {\it XXZ} \cite{ccm7,ccm8,ccm9,ccm12,ccm16}, 
{\it XY} \cite{ccm10}, and transverse Ising models \cite{ccm17}, 
and frustrated systems such as the spin-half $J_1$--$J_2$ model 
\cite{ccm13}, the spin-half triangular \cite{ccm12} and kagom\'e 
\cite{ccm18} lattice antiferromagnets, and the spin-half square-lattice
Heisenberg antiferromagnet with two types of bonds \cite{ccm},
have all been seen to be in excellent agreement with the results 
of the best of other approximate methods, where such results
exist. A recent calculation has been performed for the spin-one 
square-lattice {\it XXZ} model \cite{ccm19} which utilised the new 
``general-$s$'' formalism presented in this article, and 
the results for the ground-state properties of this model 
were found to be in superb agreement with previous SWT 
and cumulant series expansion results.

We note that although spin problems are conceptually
simple, they often demonstrate rich and unusual phase 
diagrams which are the direct result of the strong influence
of quantum fluctuations in these strongly correlated systems.
Indeed, these phase diagrams can become richer for systems with
spin quantum, $s$, greater than $1/2$, although we note that the limit 
$s \rightarrow \infty$ coincides with the pure classical model.
For example, the spin-one biquadratic model \cite{ccm3} 
contains antiferromagnetic, dimerised, and trimerised phases, and a 
``Haldane'' phase as a function of the ratio of the bond 
strengths of the linear and biquadratic terms.
Thus, lattice quantum spin problems open a wide window on to the
rapidly developing field of quantum phase transitions. This
subject becomes even more interesting when one also considers
the non-zero temperature behaviour of such systems, and the 
respective roles of quantum and thermal fluctuations in driving
phase transitions. For the present, however, we restrict ourselves
to the zero-temperature case.

\section{The Coupled Cluster Method (CCM)}

\subsection{The Ground-State Formalism}

The exact ket and bra ground-state energy 
eigenvectors, $|\Psi\rangle$ and $\langle\tilde{\Psi}|$, of a 
many-body system described by a Hamiltonian $H$, 
\begin{equation} 
H |\Psi\rangle = E_g |\Psi\rangle
\;; 
\;\;\;  
\langle\tilde{\Psi}| H = E_g \langle\tilde{\Psi}| 
\;, 
\label{eq1} 
\end{equation} 
are parametrised within the single-reference CCM as follows:   
\begin{eqnarray} 
|\Psi\rangle = {\rm e}^S |\Phi\rangle \; &;&  
\;\;\; S=\sum_{I \neq 0} {\cal S}_I C_I^{+}  \nonumber \; , \\ 
\langle\tilde{\Psi}| = \langle\Phi| \tilde{S} {\rm e}^{-S} \; &;& 
\;\;\; \tilde{S} =1 + \sum_{I \neq 0} \tilde{{\cal S}}_I C_I^{-} \; .  
\label{eq2} 
\end{eqnarray} 
The single model or reference state $|\Phi\rangle$ is normalised 
($\langle\Phi|\Phi\rangle=1$), and is required to have the 
property of being a cyclic vector with respect to two well-defined Abelian 
subalgebras of {\it multi-configurational} creation operators $\{C_I^{+}\}$ 
and their Hermitian-adjoint destruction counterparts $\{ C_I^{-} \equiv 
(C_I^{+})^\dagger \}$. Thus, $|\Phi\rangle$ plays the role of a vacuum 
state with respect to a suitable set of (mutually commuting) many-body 
creation operators $\{C_I^{+}\}$, 
\begin{equation} 
C_I^{-} |\Phi\rangle = 0 \;\; , \;\;\; I \neq 0 \; , 
\label{eq3}
\end{equation} 
with $C_0^{-} \equiv 1$, the identity operator. These operators are 
complete in the many-body Hilbert (or Fock) space,
\begin{equation} 
1=|\Phi\rangle \langle\Phi| + \sum_{I\neq 0} 
\frac {C_I^{+}  |\Phi\rangle \langle\Phi| C_I^{-}} 
{\langle\Phi| C_I^{-} C_I^{+}  |\Phi\rangle } \; . 
\label{eq4}
\end{equation} 
We note that although the 
manifest hermiticity, ($\langle \tilde{\Psi}|^\dagger = 
|\Psi\rangle/\langle\Psi|\Psi\rangle$), is lost in these 
parametrisations, the intermediate normalisation condition 
$ \langle \tilde{\Psi} | \Psi\rangle
= \langle \Phi | \Psi\rangle 
= \langle \Phi | \Phi \rangle \equiv 1$ is explicitly 
imposed. 
The {\it correlation coefficients} $\{ {\cal S}_I, \tilde{{\cal S}}_I \}$ 
are regarded as being independent variables, even though formally 
we have the relation 
\begin{equation} 
\langle \Phi| \tilde{S} =
\frac{ \langle\Phi| {\rm e}^{S^{\dagger}} {\rm e}^S } 
     { \langle\Phi| {\rm e}^{S^{\dagger}} {\rm e}^S |\Phi\rangle } \; . 
\label{eq5}
\end{equation} 
The full set $\{ {\cal S}_I, \tilde{{\cal S}}_I \}$ thus provides a complete 
description of the ground state. For instance, an arbitrary 
operator $A$ will have a ground-state expectation value given as 
\begin{equation} 
\bar{A}
\equiv \langle\tilde{\Psi}\vert A \vert\Psi\rangle
=\langle\Phi | \tilde{S} {\rm e}^{-S} A {\rm e}^S | \Phi\rangle
=\bar{A}\left( \{ {\cal S}_I,\tilde{{\cal S}}_I \} \right) 
\; .
\label{eq6}
\end{equation} 

We note that the exponentiated form of the ground-state CCM 
parametrisation of Eq. (\ref{eq2}) ensures the correct counting of 
the {\it independent} and excited correlated 
many-body clusters with respect to $|\Phi\rangle$ which are present 
in the exact ground state $|\Psi\rangle$. It also ensures the 
exact incorporation of the Goldstone linked-cluster theorem, 
which itself guarantees the size-extensivity of all relevant 
extensive physical quantities \cite{newccm8}. 

The determination of the correlation coefficients $\{ {\cal S}_I, 
\tilde{{\cal S}}_I \}$ is achieved by taking appropriate projections 
onto the ground-state 
Schr\"odinger equations of Eq. (\ref{eq1}). Equivalently, they may be 
determined variationally,
\begin{eqnarray} 
\delta{\bar{H}} / \delta{\tilde{{\cal S}}_I} =0 & \Rightarrow &   
\langle\Phi|C_I^{-} {\rm e}^{-S} H {\rm e}^S|\Phi\rangle = 0 ,  \;\; 
\forall I \neq 0 \;\; ; \label{eq7} \\ 
\delta{\bar{H}} / \delta{{\cal S}_I} =0 & \Rightarrow & 
\langle\Phi|\tilde{S} {\rm e}^{-S} [H,C_I^{+}] {\rm e}^S|\Phi\rangle 
= 0 , \;\; \forall I \neq 0 \;\; . \label{eq8}
\end{eqnarray}  
Equation (\ref{eq7}) also shows that the ground-state energy at the stationary 
point has the simple form 
\begin{equation} 
E_g = E_g ( \{{\cal S}_I\} ) = \langle\Phi| {\rm e}^{-S} H {\rm e}^S|\Phi\rangle
\;\; . 
\label{eq9}
\end{equation}  

We note that Eq. (\ref{eq7}) represents a coupled set of 
nonlinear multinomial equations for the {\it c}-number correlation 
coefficients $\{ {\cal S}_I \}$. The nested commutator expansion 
of the similarity-transformed Hamiltonian,  
\begin{equation}  
\tilde H \equiv {\rm e}^{-S} H {\rm e}^{S} = H 
+ [H,S] + {1\over2!} [[H,S],S] + \cdots 
\;\; , 
\label{eq10}
\end{equation} 
together with the fact that all of the individual components of 
$S$ in the sum in Eq. (\ref{eq2}) commute with one another, imply 
that each element of $S$ in Eq. (\ref{eq2}) is linked directly to
the Hamiltonian in each of the terms in Eq. (\ref{eq10}). Thus,
each of the coupled equations (\ref{eq7}) is of linked-cluster type.
Furthermore, each of these equations is of finite length when expanded, 
since the otherwise infinite series of Eq. (\ref{eq10}) will always 
terminate at a finite order, provided (as is usually the case) 
only that each term in the 
second-quantised form of the Hamiltonian $H$ contains a finite number of 
single-body destruction operators, defined with respect to the reference 
(vacuum) state $|\Phi\rangle$. Therefore, the CCM parametrisation naturally 
leads to a workable scheme which can be efficiently implemented 
computationally. 

The CCM formalism is exact in the limit of inclusion of
all possible multi-spin cluster correlations within 
$S$ and $\tilde S$, although in any real application 
this is usually impossible to achieve. It is therefore 
necessary to utilise various approximation schemes 
within $S$ and $\tilde{S}$. The three most commonly 
employed schemes have been: 
(1) the SUB$n$ scheme, in which all correlations 
involving only $n$ or fewer spins are retained, but no
further restriction is made concerning their spatial 
separation on the lattice; (2) the SUB$n$-$m$  
sub-approximation, in which all SUB$n$ correlations 
spanning a range of no more than $m$ contiguous lattice 
sites are retained; 
and (3) the localised LSUB$m$ scheme, in which all 
multi-spin correlations over distinct locales on the 
lattice defined by $m$ or fewer contiguous sites are 
retained. 
We also make the specific restriction that the 
creation operators $\{C_I^+\}$ in $S$ preserve
any additional symmetries of the Hamiltonian.
Thus, the approximate CCM ground-state wave function 
is constrained to lie in the appropriate subspace defined by the
additional quantum numbers corresponding to these
additional symmetries. For the {\it XXZ} model 
illustrated later the additional symmetry is provided
by the total $z$-component of spin $s_T^z = \sum_i s_i^z$,
which commutes with the Hamiltonian. The ground state lies 
in the sector $s_T^z=0$. We denote as distinct configurations
those in such appropriately defined subspace which are 
inequivalent under the point- and space-group symmetries 
of both the lattice and the Hamiltonian. The number of 
such distinct (or fundamental) configurations for the ground 
state at a given level of approximation is labelled by $N_{F}$. 

\subsection{The High-Order Formalism For General Quantum Spin Number}

In order to determine the CCM ground-state ket configurations 
we fundamentally need to {\it pattern-match} the configurations in the 
set $\{C_{I}^{-}\}$ to the spin-raising operators contained
in $\tilde H | \Phi \rangle$. We note that for small values of the 
truncation indices $\{m,n\}$ mentioned in the previous
section (and thus for low orders of approximation) this may 
readily be performed analytically. However, for higher orders 
of approximation we must use computational methods (see 
for example also Refs. \cite{ccm7,ccm8,ccm12,ccm16}) 
in order to do this. For the cases of interest here, we begin by defining 
a set of local spin axes in which all of the spins 
in the chosen model state $| \Phi \rangle$  point along 
the respective negative {\em z}-axes, namely 
\begin{equation}
\vert \Phi \rangle =\bigotimes_{i=1}^N \vert
\downarrow \rangle _i\,;
~~ \text{in the local quantization axes} , ~~
\label{eq11}
\end{equation}
where $|\downarrow \rangle _i \equiv |s,-s\rangle _i$. 
This is achieved by an appropriate set of local rotations.
Since such rotations are canonical transformations, the
underlying spin algebra is preserved, and the energy spectrum
of the transformed Hamiltonian (i.e., written in the 
rotated local spin coordinate scheme) is unchanged. 

The next step in the computational implementation of the CCM
for lattice quantum spin systems of general spin quantum 
number, $s$, is to define a suitable set of multi-spin 
creation and destruction operators with respect to this 
model state. We thus define the CCM ket-state correlation 
operator $S$ in terms of sums of products of single 
spin-raising operators, $s^+_k \equiv s^x_k + {\rm i} s^y_k$, 
(again with respect to their local spin axes), 
such that  
\begin{equation}
S= \sum_{i_1}^N {\cal S}_{i_1} s^+_{i_1} +  
\sum_{i_1,i_2}^N {\cal S}_{i_1, i_2} s^+_{i_1} s^+_{i_2} 
+ \cdot\cdot\cdot \,\, .
\label{eq12}
\end{equation}
The coefficients ${\cal S}_{i_1}$, ${\cal S}_{i_1, i_2}$, and so on,
now represent the spin-correlation coefficients specified by the sets 
of site indices, $\{i_1\}$, $\{i_1, i_2\}$ and so on, on the regular lattices 
under consideration. We note that these indices run over {\it all}
lattice sites, and that different indices may thus indicate the 
same lattice site. For the case of general spin quantum number $s$ 
we note that we have a maximum number of spin-raising operators at
any specific site $l$ which is $2 s_l$, where $s_l$ is the spin quantum 
number of the spin situated at site $l$. For the spin-half case, 
we are thus limited to only {\it one} spin-raising operator per
lattice site as required, and we note that in this manner we build 
in a previous spin-half high-order CCM formalism (see Refs. 
\cite{ccm7,ccm8,ccm12,ccm16}) into the new high-order formalism 
for general-$s$ directly from the outset. In order to simplify 
the high-order CCM formalism, it is also found to be useful to  
define the following operators:
\begin{equation}
~ \left .
\mbox{
\begin{tabular}{l@{~}l@{~}l@{~~}}
$F_{k}$   &$\equiv$ &$\sum_{l} ~ \sum_{i_2,\cdots,i_l} ~ l ~ 
{\cal S}_{k, i_2, \cdot\cdot\cdot, i_{l}} ~
s^+_{i_2} \cdot\cdot\cdot s^+_{i_{l}}$\\
$G_{k,m}$  &$\equiv$ &$\sum_{l>1} ~ \sum_{i_3,\cdots,i_l} ~ l(l-1) ~ 
{\cal S}_{k, m, i_3, \cdot\cdot\cdot, i_{l}} ~
s^+_{i_3} \cdot\cdot\cdot s^+_{i_{l}}$  \\
$M_{k,m,n}$ &$\equiv$ &$\sum_{l>2} ~ \sum_{i_4,\cdots,i_l} ~ l(l-1)(l-2) ~ 
{\cal S}_{k, m, n, i_4, \cdot\cdot\cdot, i_{l}} ~
s^+_{i_4} \cdot\cdot\cdot s^+_{i_{l}}$  \\
$N_{k,m,n,p}$&$\equiv$ &$\sum_{l>3} ~ \sum_{i_5,\cdots,i_l} ~
l(l-1)(l-2)(l-3) ~ 
{\cal S}_{k, m, n, p, i_5, \cdot\cdot\cdot, i_{l}} ~
s^+_{i_5} \cdot\cdot\cdot s^+_{i_{l}}$ 
\end{tabular}
}
\right \}
\label{eq13}
\end{equation}
Hence, we determine the similarity-transformed expressions
of the single-spin operators $s^\alpha ~ ; ~ \alpha\equiv\{+,-,z\}$,
where $s^- \equiv (s^+)^{\dag} = s^x - {\rm i} s^y$, 
by using Eq. (\ref{eq6}) and the usual spin commutation relations,
such that
\begin{equation}
~ \left .
\mbox{
\begin{tabular}{l@{~}l@{~}l@{~}l@{~~}}
$e^{-S} s_k^+ e^{S} \equiv$ &$\tilde s_k^+$  &$=$&  $s_k^+$   \\
$e^{-S} s_k^z e^{S} \equiv$ &$\tilde s_k^z$  &$=$&  $s_k^z + F_k s_k^+$   \\
$e^{-S} s_k^- e^{S} \equiv$ &$\tilde s_k^-$  &$=$&  
$s_k^- -2 F_k s_k^z - G_{kk} s_k^+- (F_k)^2 s_k^+$ ~~ .\\
\end{tabular}
}
\right \}
\label{eq14}
\end{equation} 
For the specific case of $s=1/2$ at site $k$ we note 
that $G_{kk}=0$ because ``double occupancy'' of the lattice
site $k$ is prohibited in this case. 
In order to determine the 
similarity transformed version of a given Hamiltonian, we also need 
to know the commutation relations of the operators defined in Eq. 
(\ref{eq13}) with the single-spin operators $s^\alpha ~ ; ~ 
\alpha\equiv\{+,-,z\}$, and these are stated in the Appendix. 

We now define the set of CCM destruction operators $\{C_I^-\}$ 
(of $l$ number of spin-lowering operators), as follows 
\begin{equation}
C_I^- \equiv s^-_{j_1} s^-_{j_2} \cdot\cdot\cdot ~~ s^-_{j_l} ~~ ,
\label{eq15}
\end{equation}
where the indices $j_1, j_2, \cdot\cdot\cdot, j_l$ represent
any given lattice site. 
We choose only {\it one} of the 
$N_B (l!) \nu_I$ symmetry-equivalent configurations 
to {\it pattern-match} with the terms 
within $\tilde H$ in order to determine the CCM ket-state 
equations of Eq. (\ref{eq7}). Note that $N_B$ is the number 
of Bravais lattice sites and that, for a given cluster $I$, 
$\nu_I$ is a symmetry factor dependent purely on the point-group 
symmetries (and {\it not} the translational symmetries) of 
the crystallographic lattice. 

The process of the enumeration of all 
possible fundamental clusters and the process of ``pattern-matching'' 
are both ideally suited to an efficient computational 
implementation, and a full description of these processes is 
also given in the Appendix. Furthermore, it is a simple matter 
to determine and solve the CCM bra-state equations, 
once the ket-state equations have been obtained and solved,
as described in the Appendix, where we also 
explain the technicalities involved in obtaining 
ground-state expectation values.

In the remainder of this paper we now demonstrate the new formalism
for various models of interest. 

\section{The One-Dimensional Heisenberg Model}

The 1D {\it XXZ} model is defined by the following Hamiltonian,
\begin{equation}
H ~ = ~ \sum_{i=1}^N \bigg \{ s_i^x  s_{i+1}^x +  s_i^y  s_{i+1}^y +  
\Delta s_i^z  s_{i+1}^z  \biggr \} ~~, 
\label{eq16}
\end{equation}  
where the index {\it i} in Eq. (\ref{eq16}) runs over 
all $N$ lattice sites on the linear chain. 
We are interested specifically in the infinite chain, 
$N \rightarrow \infty$. The system is by now
extremely well understood for the spin-half case via the existence
of the Bethe Ansatz \cite{ba1,ba2,ba3,ba4} exact solution for
this case. The exact ground-state energy for the infinite chain
at $\Delta=1$ was shown to be 
given by the expression, $E_g/N$=$1/4+{\rm ln}(1/2)$ ($\approx -0.44314718$). 
Indeed, the long-range N\'eel ordering inherent in
the solution in the limit $\Delta \rightarrow \infty$ is completely 
destroyed by quantum fluctuations at the phase transition point at
$\Delta=1$. By contrast, no exact solution for the spin-one 
Heisenberg model on the infinite linear chain has as yet been found,
although extremely accurate DMRG calculations have been performed 
\cite{DMRG4} for this model giving a value for the ground-state
energy of $E_g/N$=$-1.401484038971(4)$ at $\Delta=1$. 
Furthermore, these calculations conclusively showed that there is an 
excitation energy gap of magnitude $0.41050(2)$ 
in this system, which was first postulated by Haldane \cite{Haldane}.
This is in stark contrast to the spin-half model which is gapless, 
and we note that conventional spin-wave theory predicts that the
excitation energy of the Heisenberg model on the infinite linear chain 
for both the spin-half {\it and} spin-one cases is gapless. 
For the spin-one anisotropic Heisenberg model, 
the isotropic Heisenberg point is in the 
Haldane phase, in which the amount of long-range N\'eel-ordering 
is zero and the excitation spectrum is gapped. The phase transition 
from the N\'eel-like phase occurs for a value of anisotropy
$\Delta = 1.167\pm0.007$ \cite{Nomura} for the spin-one {\it XXZ} 
model.

Interest has also been expressed recently in one-dimensional ferrimagnetic 
Heisenberg systems, for which case the Lieb-Mattis theorem states 
that the system may demonstrate a macroscopic lattice (and not 
sublattice) magnetisation. For example, we note that for the 
spin-half/spin-one ferrimagnets in which alternating spins on the 
chain have, respectively, spin quantum numbers of $1/2$ and $1$, 
recent DMRG calculations \cite{DMRG5} predict that the ground-state 
energy of the Heisenberg model ($\Delta=1$) is given by 
$E_g/N \approx -0.72704$ and the {\it sublattice} magnetisation 
on the spin-one sublattice is given by $0.79248$ and the {\it 
sublattice} magnetisation (multiplied by a factor of 2 here -- see below) 
on the spin-half sublattice is given by $0.58496$.
The phase transition is also believed to be at (or near to) the 
Heisenberg point ($\Delta=1$).

The model state that we shall use in our CCM calculations here is the 
one-dimensional N\'eel state. We perform a rotation of the
local spin axes of 180$^\circ$ about the $y$-axis of the spins 
on one sublattice only, such that the model state now appears
mathematically to consist entirely of spins which points along 
the negative $z$-axis, as in Eq. (\ref{eq11}). We note however that this 
unitary transformation does not affect the eigenvalue spectrum 
of this problem. The Hamiltonian may now be written in terms 
of these local spin axes, as
\begin{equation}
H ~ = ~ - \sum_{i=1}^N \biggl \{ 
\Delta s_i^z s_{i+1}^z + \frac 12 
\biggl (s_i^+ s_{i+1}^+ + s_i^- s_{i+1}^-\biggr )
\biggr \} ~~, 
\label{eq17}
\end{equation}  
where $s^{\pm} = s^x \pm {\rm i} s^y$. For the spin-half and
spin-one antiferromagnets, we define the sublattice magnetisation
in terms of these local spin-axes as,
 \begin{equation}
M = - \frac 1{sN} \sum_{i=1}^N \langle \tilde \Psi | s_{i}^z | \Psi  
\rangle  ~~ ,
\label{eq18}
\end{equation}
where the index $i$ runs over all $N$ lattice sites. This reflects
the fact that for the spin-half and spin-one Heisenberg models 
we have a unit cell which contains one single lattice site only. 
By contrast, for the spin-half/spin-one ferrimagnet we have a 
unit cell which contains two nearest-neighbour spins of quantum 
spin numbers $1/2$ and $1$. We thus define the sublattice 
magnetisations $M_1$, on the $A$ (spin-half) sublattice 
sites, and $M_2$, on the $B$ (spin-one) sublattice sites, 
separately, in the local spin axes by
\begin{eqnarray}
M_1 &=& -\frac 2{N_1} \sum_{i_1}^{N_1} \langle \tilde \Psi | s_{i_1}^z  | \Psi  
\rangle ~~ {\rm and} \label{eq19} \\
M_2 &=& -\frac 1{N_2} \sum_{i_2}^{N_2} \langle \tilde \Psi | s_{i_2}^z | \Psi  
\rangle ~~  , \label{eq20} 
\end{eqnarray}
where $i_1$ runs over all $N_1=N/2$ spin-half sites and $i_2$ runs 
over all $N_2=N/2$ spin-one sites. Furthermore, we note that the 
Lieb-Mattis theorem states that $|M_1/2-M_2|=1/2$ (after rotation of 
the local spin axes) thus allowing an overall macroscopic magnetic 
moment. The technicalities of determining $M$, $M_1$, and $M_2$
are explained in the Appendix.

We may now perform the similarity transform of the Hamiltonian
in order to write that Hamiltonian, in terms of the operators 
$F_{k}$, $G_{k,m}$, $M_{k,m,n}$, and $N_{k,m,n,p}$ given above. 
By making use of Eqs. (\ref{eq14}) and (\ref{appendix9}) we find that 
$\tilde H | \Phi \rangle \equiv e^{-S} H e^S | \Phi \rangle =
(\tilde H_1 + \tilde H_2 + \tilde H_3) | \Phi \rangle$, where
\begin{eqnarray} 
\tilde H_1 &\equiv&
- \sum_{i}
\biggl\{
\Delta \biggl (G_{i,{i+1}}+ F_i F_{i+1} \biggr ) + 
\frac 12 
\biggl (1 + 
N_{i,i,i+1,i+1}+
2 G^2_{i,i+1}+
2 M_{i,i,i+1} F_{i+1} +
2 F_{i} M_{i,i+1,i+1} 
\nonumber \\
&& 
+ 4 F_{i} F_{i+1} G_{i,i+1}+
G_{i,i} G_{i+1,i+1} +
G_{i,i} F_{i+1}^2 +
F_{i}^2 G_{i+1,i+1} +
F_{i}^2 F_{i+1}^2 \biggr ) ~ \biggr\} ~
s_i^+ s_{i+1}^+ ~ ,
\label{eq21a}   \\
\tilde H_2 &\equiv&
-\sum_{i} 
\biggl\{
\biggl (
\Delta F_i +
M_{i,i,i+1} +
2 F_i G_{i,i+1} +
G_{i,i} F_{i+1} +
F_i^2 F_{i+1} 
\biggr ) 
s_i^+ s_{i+1}^z 
\nonumber \\
&& 
+ \biggl (
\Delta F_{i+1} +
M_{i,i+1,i+1} +
2 G_{i,i+1} F_{i+1} +
F_i G_{i+1,i+1} +
F_i F_{i+1}^2
\biggr ) 
s_{i}^z s_{i+1}^+ ~~  
\biggr\} ~~ ,
\label{eq21b} \\ 
\tilde H_3 &\equiv&
-\sum_{i} 
\biggl \{ 
\Delta + 2G_{i,{i+1}}+ 2F_i F_{i+1}
\biggr\} ~ 
s_i^z s_{i+1}^z
~~ .
\label{eq21c}
\end{eqnarray}
Again, we note that repeated indices in the operators
$G_{k,m}$, $M_{k,m,n}$, and $N_{k,m,n,p}$ operators are prohibited 
for the spin-half case. Hence, the general spin quantum number 
formalism reduces to the previous formalism determined for 
the spin-half case in this limit \cite{ccm7,ccm8,ccm12,ccm16}. 
A general feature of CCM calculations for spin systems is that 
we often solve the CCM at different values of some tunable 
parameter within the Hamiltonian. For the case of the 
Hamiltonian of Eq. (\ref{eq17}) this is the anisotropy parameter, 
$\Delta$, and we note that we ``track'' the solution 
from the trivial limit $\Delta \rightarrow \infty$ to lower 
values of $\Delta$. Another feature of CCM calculations is 
that often the real solution to the CCM equations is seen 
to terminate at some critical value of this parameter 
(e.g., denoted $\Delta_c$ here), and this behaviour is taken 
to be a signal of a phase transition in the real system. This 
illustrates a strong advantage of this method: namely, that the
CCM is able to make predictions for the positions of quantum phase 
transition points from within a fully {\it ab initio} framework.

\section{Results}

In this article we utilise two approximation schemes 
in order to determine CCM expectation values for the
ground-state energy and the sublattice magnetisation.
The latter gives a measure of the amount of sublattice ordering. 
The two schemes are, namely, the LSUB$m$ and SUB$m$-$m$ 
approximation schemes, both of which include correlations 
in a systematic and structured manner and provide exact 
results in the asymptotic limit $m \rightarrow 
\infty$. We note that the LSUB$m$ and SUB$m$-$m$ schemes 
are equivalent for the ferrimagnet, if the truncation
index $m$ is an even number. This is due to 
the restriction that $s_T^z= \sum_i s_i^z =0$ in the 
ground state thus ruling out any clusters in the LSUB$m$
approximation which contain more than $m$ spin flips.
This is {\it not} the case for the spin-one antiferromagnet,
in which case the restriction $s_T^z= \sum_i s_i^z =0$ 
may be satisfied for various clusters in LSUB$m$ which
do contain more than $m$ spin flips.

However, we note that it is usually impossible in
a practical application of this method to exactly solve
for the LSUB$m$ and SUB$m$-$m$ schemes in the limit 
$m \rightarrow \infty$, and so we extrapolate the ``raw'' 
results as a function of $m$ in this limit. Although
no rigorous extrapolation scaling laws exist, we
may attempt to extrapolate these results ``heuristically.'' 
Thus, for example, extrapolation of these 
results in the limit $m \rightarrow \infty$ at each value 
of $\Delta$ independently may be achieved by assuming a 
``power-law'' dependence, hereforth denoted as the 
``extrapolated CCM'' results, given by
\begin{equation}
y_i = a + b x_i^{\nu} ~~, 
\end{equation}
where $y_i$ is a CCM expectation value at a given level of 
LSUB$m$ or SUB$m$-$m$ approximation and $x_i = 1/m$, and there 
are $p$ data points (i.e., for the $p$ different values of
$m$ for which calculations have been performed). The {\it 
best fit} of this data to the parametric form given in the
equation above is found and the extrapolated result is thus
given by $a$. The interested reader will find a comprehensive 
explanation of the extrapolation process of CCM LSUB$m$ 
expectation values for the spin-half {\it XXZ} model for a 
variety of lattices in Refs. \cite{ccm16,ccm17}.

Results for the {\it XXZ} model are presented in 
Figs. \ref{fig1}--\ref{fig4}, and results for the Heisenberg 
model ($\Delta=1$) on the linear chain are given in Tables \ref{tab1}
--\ref{tab3}. 
We may see from  Figs. \ref{fig1} and \ref{fig_energy_spin_one} for 
the spin-half and spin-one {\it XXZ} models that our results for the
ground-state energy are highly converged, and analogous
behaviour is seen for the ground-state energies of 
the spin-half/spin-one ferrimagnet. 
Indeed, we may see from Tables \ref{tab1}--\ref{tab3} 
that our CCM results for the Heisenberg model ($\Delta=1$) 
converge rapidly in each case, and that they agree well with 
their respective exact or DMRG results. 
We note that the extrapolated CCM results
for the spin-one Heisenberg model seem to agree 
with DMRG results to only two decimal places, whereas 
the extrapolated CCM results for the spin-half 
Heisenberg model appear to agree with exact Bethe
Ansatz results to about five decimal places.
The apparent discrepancy is understood by 
noting that we are well into the Haldane 
phase at $\Delta=1$ for the spin-one model
and the fact that we obtain results at this
point at all is a testament to the power of
the CCM. We note however that other model
states might be employed at the Heisenberg
point for the spin-one model in order to 
obtain even more accurate results for this model.
Inspection of Table \ref{tab3} (for $\Delta=1$) 
indicate that our ``raw'' (unextrapolated) CCM 
results for the ground-state energy of the spin-half/spin-one 
ferrimagnet appear already to be converged to {\it at least} 
five decimal places. We expect that the CCM 
results are even more accurate for $\Delta > 1$
because quantum fluctuations are (comparatively)
weaker  in this regime than at the isotropic Heisenberg 
point at $\Delta=1$. 
We furthermore expect that our extrapolated CCM value 
at $\Delta=1$ given in Table \ref{tab3} represents an 
even better result than the raw SUB$m$-$m$ results.
Indeed, we note that all of our CCM results are in 
excellent agreement with those result of the DMRG 
method \cite{DMRG5} at this point.

Figs. \ref{fig2} and \ref{fig3} present 
results for the sublattice magnetisation, $M$, 
for the spin-half and  spin-one {\it XXZ} models 
in the N\'eel-ordered regime. 
The extrapolated CCM results for the spin-half {\it XXZ} model 
are clearly in excellent agreement with exact results and we
note in particular that the sublattice magnetisation goes to
zero at $\Delta=1.07$. This is in good agreement with
the exact behaviour of the model and we note that there
is an infinite-order phase transition at $\Delta = 1$
to disordered phase for $\Delta \le 1$ in the `real' system.
Similarly, the extrapolated CCM results for both the SUB$m$-$m$ and 
LSUB$m$ approximation schemes for the spin-one {\it XXZ} model 
go to zero at $\Delta \approx 1.19$ and $\Delta \approx 1.18$, 
respectively. This is in excellent agreement with the position for the
onset of the Haldane phase of $\Delta = 1.167\pm0.007$
\cite{Nomura}.
We furthermore note that even greater accuracy for the CCM 
extrapolated results for both the spin-half and spin-one
antiferromagnets is expected with higher levels of 
approximation. We also note that these results are a clear 
indication that the CCM is capturing the qualitative
difference in the behaviour of the sublattice magnetisations
between the spin-half and spin-one antiferromagnetic \
{\it XXZ} models. 

Furthermore, an overall magnetic moment is possible
for the case of the spin-half/spin-one ferrimagnet and we may 
see from Fig. \ref{fig4} and Table \ref{tab3} that the CCM 
results for the amount of sublattice ordering on both sublattices,
namely, $M_1$ and $M_2$, are extremely well converged over the 
whole of the N\'eel-like regime. Indeed, we note that 
even though the values for $M_1$ and $M_2$ are determined 
separately using the CCM, they still obey relationship 
$|M_1/2-M_2|=1/2$ specified by the Lieb-Mattis theorem.
It is furthermore not {\it a priori} evident that this 
ought to be the case as we have not artificially
constrained this to be true in our CCM calculation 
at any point. It is thus a reflection that the CCM is 
detecting the underlying nature of the ground state from
within an {\it ab initio} framework. We also see from 
Table \ref{tab3} that the CCM results agree
with DMRG results to (at least) about five decimal places
for the isotropic model. Furthermore, we believe
that the extrapolated CCM results for the sublattice
magnetisations at the Heisenberg point at $\Delta=1$ 
present even more accurate estimates of these quantities. 
We note that the results for the sublattice magnetisations
of the spin-half/spin-one ferrimagnet remain non-zero 
and finite over the whole of the regime $\Delta \ge 1$,
although it is possible that a phase transition to a regime 
in which spins lie in the $xy$-plane (and for which 
$M_1=M_2=0$) occurs for $\Delta \approx 1$ (for example, see
Refs. \cite{ccm9,ccm12}). (We note that we would probably employ a model 
state in which spin lie in the $xy$-plane for the CCM in 
the region $\Delta \le 1$.) Our highly converged 
CCM results thus indicate that if the phase transition does 
indeed occur near to $\Delta \approx 1$ then this might
involve a stepwise change in the sublattice magnetisations from a 
non-zero (and finite) value to zero at this point. If this 
conjecture were true then this behaviour would furthermore 
be consistent with a first-order phase transition at this
point.

Finally, we note that the CCM critical points, $\Delta_c$, 
are observed for the spin-one antiferromagnetic {\it XXZ}
model (illustrated in Table \ref{tab2}) at which the real (i.e., physical) 
solution to the CCM equations breaks down at given level of LSUB$m$
or SUB$m$-$m$ approximation level. This behaviour is an indication
that a phase transition occurs in the ``real'' system. 
We note that an alternative and independent indication of the 
position of the phase transition is afforded by the value of
$\Delta_c$ at which the sublattice magnetisation  
goes to zero (discussed above). No equivalent results for the 
critical points of the ferrimagnetic {\it XXZ} model were obtained, 
although CCM calculations for this model show characteristic 
peaks in the derivatives of the ground-state energy for $\Delta 
\le 1$. This behaviour might possibly again be interpreted
as a signature of a phase transition. It is also quite possible 
however that this typical breakdown of the real solution of the CCM 
equations might however occur at higher orders of approximation 
\cite{ccm9}. 

\section{Conclusions}

In this article a high-order CCM ground-state formalism for 
lattices comprised of spins with general spin 
quantum number $s$ was presented. We have shown that this new 
formalism is highly suitable for computational implementation for 
localised approximation schemes (namely, the LSUB$m$ and the SUB$m$-$m$
approximation schemes).
In order to demonstrate 
the effectiveness of the formalism, successful applications of the 
general-$s$ formalism were made to the 
one-dimensional {\it XXZ} model for the spin-half and spin-one 
antiferromagnetic cases and the spin-half/spin-one ferrimagnetic case. 
Our extrapolated results were seen to agree extremely well with exact 
results for the spin-half antiferromagnetic {\it XXZ} model, up to 
and including the phase transition point at $\Delta=1$. 
Our extrapolated results for the spin-one antiferromagnetic {\it XXZ} 
model predicted that the sublattice magnetisation goes to 
zero at $\Delta \approx 1.2$. This result is in 
excellent agreement with the value for the onset of the Haldane phase 
of $\Delta = 1.167\pm0.007$ \cite{Nomura} 
for this model. Finally, CCM results for the sublattice magnetisation 
(on both sublattices) of the spin-half/spin-one ferrimagnet 
were seen to be extremely well converged as a function of the
truncation index $m$, and these results were finite and non-zero 
over a wide range of $\Delta$, up to and including the 
Heisenberg point at $\Delta=1$. 


Further applications of the new high-order general-$s$ CCM formalism
to the zero-temperature properties of lattice quantum spin systems
are also envisaged for the future. In particular, it is expected
that such techniques will be applied to highly frustrated cases 
with spatial dimensionality greater than one which are difficult 
(if not impossible) to treat using other approximate theories. 
It is also straightforward to extend the ground-state formalism
presented here to deal with excited states, as has been 
done previously for the spin-half case \cite{ccm16,ccm}.

\section*{Acknowledgements}

We thank Dr. Chen Zeng for his useful and interesting 
discussions. One of us (KAG) gratefully acknowledges a research grant 
(GR/M45429) from the Engineering and Physical Sciences Research Council 
(EPSRC) of Great Britain. This work has also been supported by the Deutsche
Forschungsgemeinschaft (GRK 549, Graduiertenkolleg on `Accentric Crystals'
and also Project No. RI 615/9-1).

\appendix

\section{}

\subsection{Commutation Relations and The High-Order General-$s$ CCM Formalism}

In this article we present a new formalism and results for high-order
ground-state CCM calculations for general spin quantum number, $s$, 
based on a model state in which all spins on the crystallographic lattice
point downwards along the local $z$-axes. A large part of the new 
formalism relies on the new ``high-order'' CCM operators defined by
Eq. (\ref{eq13}) and also their commutation relations with the single-spin
operators in order to determine the similarity transforms of
various operators, such as the Hamiltonian for example. 
In order to determine these commutation relations we firstly 
remind ourselves that the ket-state correlation operator $S$ is given 
by Eq. (\ref{eq2}) with 
$C_I^+ \equiv  s^+_{i_1} s^+_{i_2} \cdot\cdot\cdot s^+_{i_l}$
and ${\cal S}_I \equiv  {\cal S}_{i_1,i_2,\cdot\cdot\cdot,i_l}$,
and hence
\begin{equation}
S = \sum_l \sum_{i_1,i_2,\cdot\cdot\cdot,i_l} 
{\cal S}_{i_1,i_2,\cdot\cdot\cdot,i_l} ~ 
s^+_{i_1} s^+_{i_2} \cdot\cdot\cdot s^+_{i_l}
\label{appendix2} ~~ ,
\end{equation}
where each of the indices $\{i_1,i_2,\cdot\cdot\cdot,i_l\}$ runs over all
lattice sites with the condition that there can be no more than
$2s$ of them at any particular lattice site. The usual spin commutation 
relations of the spin operators also apply, 
\begin{eqnarray} 
[s_l^+,s_{l'}^-] = 2 s_l^z \delta_{l,l'} ~~  &;&  
~~ [s_l^z,s_{l'}^{\pm}] = \pm s_l^{\pm} \delta_{l,l'} \label{appendix3} ~~ . 
\end{eqnarray} 

We also note that the commutation of a given operator with $S$ must
be distributive, such that
\begin{eqnarray} 
[s_k^{\alpha},S] ~ &=& ~  
\sum_l \sum_{i_1,i_2,\cdot\cdot\cdot,i_l} 
{\cal S}_{i_1,i_2,\cdot\cdot\cdot,i_l}
\biggl \{ [s_k^{\alpha}, s_{i_1}^+] s^+_{i_2} \cdot\cdot\cdot s^+_{i_l} +
s_{i_1}^+ [s_k^{\alpha}, s_{i_2}^+] s^+_{i_3} \cdot\cdot\cdot s^+_{i_l} 
\nonumber  \\
&& ~~~~~~~~~~~~ + ~ \cdots ~ + s_{i_1}^+ s_{i_2}^+  
\cdot\cdot\cdot [s_k^{\alpha}, s_{i_l}^+]
\biggr \} ~~ ,
\label{appendix4}
\end{eqnarray} 
where $\alpha=\{z,+,-\}$. As pairs of spin-raising operators always 
commute, we may therefore state that $[s_k^{+},S]=0$. Furthermore, 
for the case of $[s_k^{z},S]$ we note again that each index runs 
over all lattice sites, which implies that each term on the right-hand 
side of Eq. (\ref{appendix4}) is equivalent and that, as there 
are $l$ such terms, we may write this expression as
\begin{equation}
[s_k^{z},S]  = \sum_l \sum_{i_2,\cdot\cdot\cdot,i_l} l ~ 
{\cal S}_{k,i_2,\cdot\cdot\cdot,i_l} 
s^+_{i_2} \cdot\cdot\cdot s^+_{i_l} s_k^+ = F_k s_k^+
\label{appendix5} ~~ .
\end{equation}
Note again that the ``high-order'' operators such as $F_k$ are defined by 
Eq. (\ref{eq13}). We lastly calculate the commutator $[s_k^-,S]$
in Eq. (\ref{appendix4}), and 
using the basic commutation relations of Eq. (\ref{appendix3})
we thus have
\begin{eqnarray} 
[s_k^{-},S] ~ &=& ~  
-2 \sum_l \sum_{i_1,i_2,\cdot\cdot\cdot,i_l} 
{\cal S}_{i_1,i_2,\cdot\cdot\cdot,i_l}
\biggl \{  \delta_{k,i_1} s_k^{z} s^+_{i_2} \cdot\cdot\cdot s^+_{i_l} +
\delta_{k,i_2} s_{i_1}^+ s_k^z s^+_{i_3} \cdot\cdot\cdot s^+_{i_l} 
\nonumber  \\
&& ~~~~~~~~~~~~ + ~ \cdots ~ + \delta_{k,i_l} s_{i_1}^+ s_{i_2}^+  
\cdot\cdot\cdot s_k^z
\biggr \} ~~ ,
\label{appendix6}
\end{eqnarray} 
We now commute the operator $s_k^z$ past the strings of 
spin-raising operators in Eq. (\ref{appendix6}) using the
basic commutation relations of Eq. (\ref{appendix3}). Thus,
for example,
\begin{eqnarray}
s_k^{z} s^+_{i_2} s^+_{i_3} \cdot\cdot\cdot s^+_{i_l} &=&
(\delta_{k,i_2} + \delta_{k,i_3} + ~ \cdot\cdot\cdot ~
\delta_{k,i_l}) s^+_{i_2} s^+_{i_3} \cdot\cdot\cdot s^+_{i_l} 
\nonumber \\
&&~~~~~~~ +  ~ s^+_{i_2} s^+_{i_3} \cdot\cdot\cdot s^+_{i_l} s_k^{z}
\label{appendix7}
\end{eqnarray}
By inserting Eq. (\ref{appendix7}) into Eq. (\ref{appendix6})
we find that
\begin{eqnarray} 
[s_k^{-},S] ~ &=& ~ -2 \sum_l \sum_{i_3,i_4,\cdot\cdot\cdot,i_l} 
\bigl ( \sum_{n=1}^{l-1} \bigr ) {\cal S}_{k,k,i_3,\cdot\cdot\cdot,i_l} ~
 s^+_{i_3} s^+_{i_4} \cdot\cdot\cdot s^+_{i_l} s^+_{k} 
\nonumber  \\
&& ~~~~~~ -2 \sum_l \sum_{i_2,i_3,\cdot\cdot\cdot,i_l} 
l {\cal S}_{k,i_2,\cdot\cdot\cdot,i_l} ~  
s_{i_2}^+ s_{i_3}^+  \cdot\cdot\cdot s^+_{i_l} s^z_{k} 
\nonumber  \\
&=& -G_{k,k} s_k^+ - 2F_k s_k^z  
~~ ,
\label{appendix8}
\end{eqnarray} 
using the definitions in Eq. (\ref{eq13}). We note again that
for the case $s=1/2$ the operator $G_{k,k} \equiv 0$. 

By making use of the nested commutator expansion for the 
similarity-transformed operators [c.f., Eq. (\ref{eq10})],
it is now a simple matter to verify the relations in Eq. (\ref{eq14}),
by using Eqs. (\ref{appendix5}) and (\ref{appendix8}).

In order to determine the similarity transform of the Hamiltonian 
however it  is also necessary to also know the commutation relations 
of the single-spin operators with respect to $F_k$, $F_k^2$, $G_{k,m}$,
and $M_{k,m,n}$. The proofs of these commutation relations follow a 
similar pattern to the proofs given above, and so we merely state them here:
\begin{equation}
~ \left .
\mbox{
\begin{tabular}{l@{~}l@{~}l@{~~}}
$[s_k^{z},F_{m}]$    ~ &=& ~ $G_{k,m} s_k^+$~~,\\
$[s_k^{z},G_{m,n}]$  ~ &=& ~ $M_{k,m,n} s_k^+$~~,\\
$[s_k^{z},F_m^2]$    ~ &=& ~ $2 F_m G_{k,m} s_k^+$~~,\\
$[s_k^{-},F_{m}]$    ~ &=& ~ $-2 G_{k,m} s_k^z - M_{k,k,m} s_k^+$~~,\\
$[s_k^{-},F_{m}^2]$  ~ &=& ~ $-2 G_{k,m}^2 s_k^+ - 2 F_m M_{k,k,m} s_k^+
-4F_m G_{k,m} s_k^z $~~,\\
$[s_k^{z},M_{m,n,p}]$  ~ &=& ~ $N_{k,m,n,p} s_k^+$~~,\\
$[s_k^{-},G_{m,n}]$  ~ &=& ~ $-2 M_{k,m,n} s_k^z - N_{k,k,m,n} s_k^+$~~.\\
\end{tabular}
}
\right \} ~~
\label{appendix9}
\end{equation}
We note once more that the operators $F_k$, $G_{k,m}$, 
$M_{k,m,n}$, and $N_{k,m,n,p}$ are defined by Eq. 
(\ref{eq13}). 

\subsection{Enumeration of the Fundamental Clusters}

At a given level of approximation, we choose only {\it one} of 
the $N_B (l!) \nu_I$ possible symmetry-equivalent configurations 
for a given {\it fundamental} configuration of $l$ spin-raising
operators, where $N_B$ is 
the total number of Bravais lattice sites, and where $\nu_I$ 
is a symmetry factor dependent purely on the point-group 
symmetries (and {\it not} the translational symmetries) 
for the crystallographic lattice in question and for 
fundamental configuration $I$. We note that there are $N_F$ 
such {\it fundamental} configurations. 
The first part of the computational algorithm is to 
enumerate all of the ``lattice animals'' which define the
``locale'' in which the clusters must lie. For the
levels of approximation shown in this article it is possible
to do this by using a simple recursive algorithm which
enumerates all possible lattice animals of $m$ contiguous
sites. This ``locale'' is explicitly assumed here to be the same for 
both the LSUB$m$ and SUB$n$-$m$ approximation schemes. 
Secondly, one then needs to enumerate all possible ways in which one 
can place ($2s$) or less spin-raising operators on each of the positions 
of the $m$ sites within each of these lattice animals. There are 
thus $(2s)^m$ possibilities for each lattice animal. However,
one must also restrict the total number of spin-raising operators 
to be less than or equal to $n$ for the SUB$n$-$m$ approximation 
scheme. We note however that there is no such restriction on 
the total number of spin-raising operators for the LSUB$m$ approximation. 
This process thus enumerates all possible connected and 
disconnected clusters, and we make a restriction that we
include only those clusters which are inequivalent under the 
point and space group symmetries of both the lattice and the 
Hamiltonian. A further restriction for the systems under consideration
in this article is that we must restrict the set of fundamental clusters 
to include only those which preserve the relationship, 
$s^z_T=\sum_i s^z_i=0$, with respect to the original (``unrotated'') 
N\'eel model state since $[s^z_T,H]=0$ and the ground state
lies in the $s^z_T=0$ sector.

\subsection{The Ket-State Equations}

We now wish to determine the CCM ket-state equations, where the 
$I$-th such equation is given by
\begin{equation}
E_I \equiv \frac 1{A_I} \langle \Phi | C_I^- e^{-S} H e^{S} | \Phi \rangle = 0 ~~,
\forall I \ne 0 ~~ ,
\label{tempLabel}
\end{equation}
where $A_I$ is a normalisation factor given by
$A_I \equiv \langle \Phi | C_I^- C_I^+ | \Phi \rangle$
$=  \langle \Phi | (s_{i_1}^- s_{i_2}^- \cdots s_{i_l}^- ) ~
(s_{i_1}^+ s_{i_2}^+ \cdots s_{i_l}^+ ) | \Phi \rangle$. 
We note once more that we choose only {\it one} of the 
$N_B (l!) \nu_I$ possible symmetry-equivalent configurations 
for a given {\it fundamental} configuration in $C_I^-$ in order 
to pattern-match with the terms within $\tilde H | \Phi \rangle$ 
and thus determine the $I$-th CCM ket-state equation.
We then computationally match the individual 
spin-lowering operators in $C_I^-$, defined by Eq. (\ref{eq15}), 
to the spin-raising operators in $\tilde H | \Phi \rangle$.
We therefore put constraints on the indices in the CCM ket-state
correlation coefficients, $\{ {\cal S}_{i_1,i_2,\cdots,i_l}\}$,
and these constraints on the indices allow us to enumerate all 
possible terms which contribute to the CCM ket-state equations.
For example, we may consider a specific term in 
the evaluation of the CCM ket-state equations, given by 
\begin{eqnarray}
\langle \Phi | C_I^- ~ \tilde s^z_k \tilde s^z_m | \Phi \rangle &=& 
\langle \Phi | C_I^- ~ \biggl ( 
s^z_k  s^z_m + F_m s_m^+ s_k^z + F_k s_k^+ s_m^z \nonumber \\
&& ~~~~~~~~~~~~~~~~~~~~~~~ + 
G_{k,m} s_k^+ s_m^+ + F_k F_m  s_k^+ s_m^+ \biggr ) 
| \Phi \rangle ~~ .
\label{appendix_2}
\end{eqnarray}
For the case of the linear chain Heisenberg model we 
let $k$ run over all lattice sites on the 1D chain and we set 
$m=i+1$. Now consider a specific term within Eq. (\ref{appendix_2}), 
given by $\langle \Phi | C_I^- F_k F_m  s_k^+ s_m^+ 
| \Phi \rangle$. We match the indices of the spin-lowering
operators in $C_I^-$ to the spin raising-operators in $F_k$,
$F_m$, $s_k^+$, and $s_m^+$. Hence, both $k$ and $m$ are fully
constrained to take on site values dependent on those indices
of the spin-lowering operators in the fundamental configuration
chosen for $C_I^-$. We note however that one may not always have 
such complete constraints on the indices $k$ and $m$ in $\tilde H$. For 
example, we may attempt to evaluate such a term as, $\langle 
\Phi | C_I^- F_k F_m  s_k^z s_m^z | \Phi \rangle$. In this case, 
both $k$ and $m$ are free to cover all lattice sites {\it independently} 
from the fundamental cluster utilised in $C_I^-$. However, 
we may retain only those configurations in the set $\{ {\cal 
S}_{i_1,i_2,\cdots,i_l} \}$ in $F_k$ and $F_m$ which are equivalent 
to the fundamental set of configurations under the symmetries 
of the lattice. This condition is sufficient to render the 
computational problem both tractable and efficient. Finally, 
the resulting coupled, non-linear CCM ket-state equations are 
easily solved computationally (for example, via the Newton-Raphson
method) at a given value of the anisotropy parameter $\Delta$.

\subsection{The Bra-State Equations}

The bra-state coefficients ${{\cal S}_I}$ of Eq. (\ref{eq2}) 
are formally determined by Eq. (\ref{eq8}). 
However, this form of the bra-state equations is slightly cumbersome to 
use, and a simpler and more elegant approach is possible by defining the 
following {\it new} set of CCM correlation coefficients given by
\begin{equation}
~ \left .
\mbox{
\begin{tabular}{l@{~}l@{~}l@{~~}}
$x_I$ &=& ${\cal S}_I$  \\
$\tilde x_I$ &=& $\frac {N_B}N  
\tilde {\cal S}_I  A_I  \nu_I  (l!)$  \\
\end{tabular}
}
\right \} ~~ ,
\label{appendix14}
\end{equation}
where again $A_I \equiv \langle \Phi | C_I^- C_I^+ | \Phi \rangle$
$=  \langle \Phi | (s_{i_1}^- s_{i_2}^- \cdots s_{i_l}^- ) ~
(s_{i_1}^+ s_{i_2}^+ \cdots s_{i_l}^+ ) | \Phi \rangle$. 
Note that $N_B$ is the number of Bravais lattice sites. Note also 
that for a given cluster $I$ then $\nu_I$ is a symmetry factor which 
is dependent purely on the point-group symmetries (and {\it not} 
the translational symmetries) of the crystallographic lattice and 
that $l$ is the number of spin operators. We note 
that the coefficients $A_I$, $\nu_I$, and $N_B$ however do not need 
to be explicitly determined because they {\it always} cancel out 
when obtaining ground-state expectation values (see below). 
The CCM bra-state operator may thus be rewritten as
\begin{equation}
\tilde S \equiv 1 + N \sum_{I=1}^{N_F} \frac{\tilde x_I}{A_I} C_I^- ~~ ,
\end{equation}
such that
\begin{equation}
\bar{H} = E_0 + N \sum_I^{N_F} \tilde x_I E_I ~~.
\label{appendix15}
\end{equation}
We note again that the ground-state energy expectation value is
defined by $E_0 = \langle \Phi| e^{-S} H e^S | \Phi \rangle$ 
and that $E_I$ is the $I$-th CCM ket-state equation defined
by Eq. (\ref{tempLabel}). The CCM ket-state equations are easily 
rederived by taking the partial derivative of $\bar{H}/N$ with 
respect to $\tilde x_I$, where
\begin{equation}
0 = \frac {\delta{(\bar{H}/N)}}{\delta \tilde x_I} \equiv E_I ~~.
\label{appendix16}
\end{equation}
We now take the partial derivative of $\bar{H}/N$ with respect to 
$x_I$ such that the bra-state equations take on a particularly 
simple form, given by
\begin{equation}
0 = \frac {\delta{(\bar{H}/N)}}{\delta x_I} = 
\frac {\delta{(E_0/N)}}{\delta x_I} + 
\sum_{J=1}^{N_F} \tilde x_J \frac {\delta{E_J}}{\delta x_I} ~~.
\label{appendix17}
\end{equation}
This equation is easily solved computationally, once the CCM ket-state
equations have been determined and solved, and the numerical values 
of the coefficients $\{\tilde x_I \}$ may thus be obtained. We note
that this approach greatly simplifies the task of determining the
bra-state equations because we never need to explicitly determine
the factors $N_B$, $A_I$, or $\nu_I$. 

\subsection{Expectation Values}

Expectation values of spin operators may be treated in an analogous
manner to that of the expectation value of the Hamiltonian, given by 
$\bar{H}$. For example, the sublattice magnetisation for the 
spin-half and spin-one antiferromagnets of Eq. (\ref{eq18}) 
may be written as
\begin{eqnarray}
M &=& - \frac 1{sN} \sum_{i=1}^N \langle \tilde \Psi | s_{i}^z | \Psi  
\rangle  ~~  \nonumber \\
&=& 1 - \frac {1}{s} \sum_{I=1}^{N_F} \frac{\tilde x_I}{A_I} \langle \Phi 
| C_I^- \sum_{i=1}^N (F_i s_i^+) | \Phi \rangle  ~~  \nonumber \\
&=& 1 - \frac 1s \sum_{I=1}^{N_F} ~ l (l!) ~ \tilde x_I x_I  ~~ ,
\label{appendix18}
\end{eqnarray}
where $i$ runs over all lattice sites. We again note 
that the factors $A_I$ and $\nu_I$ in Eq. ({\ref{appendix18})
have cancelled out. Equation ({\ref{appendix18}) is easily evaluated 
once the ket- and bra-state equations have been solved at a given value 
of the anisotropy parameter $\Delta$. (Results for both the ground-state 
energy and the sublattice magnetisation are given in Section III of this 
article.) 

The situation for the ferrimagnet is slightly different, because
the unit cell now contains two spins. Thus the magnetisation 
on $N_1$ the spin-half sites ($s=1/2$) is given by,
\begin{eqnarray}
M_{1} &=& -\frac 1{s N_{1}} \sum_{i_{1}}^{N_{1}} 
\langle \tilde \Psi | s_{i_1}^z  | \Psi  
\rangle ~~  \nonumber \\
&=& 1 - \frac {N}{sN_1} \sum_{I=1}^{N_F} \frac{\tilde x_I}{A_I} 
\langle \Phi | C_I^- \sum_{i_1=1}^{N_1} (F_{i_1} s_{i_1}^+) | \Phi \rangle  ~~ 
\nonumber \\
&=& 1 - 4 \sum_{I=1}^{N_F} \frac{\tilde x_I}{A_I} \langle \Phi 
| C_I^- \sum_{i_1=1}^{N_1} (F_{i_1} s_{i_1}^+) | \Phi \rangle ~~ .
\end{eqnarray}
Note that ${i_{1}}$ runs over all $N_1(=N/2)$ spin-half 
lattice sites, such that we have the factor $\frac {N}{sN_1}=4$. 
It is a simple matter to explicitly enumerate all combinations of 
orderings of the $l$ spin-raising operators in $\sum_{i_1=1}^{N_1} 
F_{i_1} s_{i_1}^+$ which match with spin-lowering operators 
in $C_I^-$, although we must also explicitly restrict $i_1$ to 
be a spin-half site. (We note that the factors $A_I$ again cancel out 
with coefficients in $\langle \Phi | C_I^- \sum_{i_1=1}^{N_1} (F_{i_1} 
s_{i_1}^+) | \Phi \rangle$ at this point.) 
A similar expression may be obtained for $M_2$.

We may also determine numerical values for other expectation 
values, such as the spin-spin correlation function, given by
\begin{eqnarray}
M_{r}^{zz} &\equiv& \frac 1N \sum_{k=1}^N \langle \tilde \Psi |
\tilde s^z_k \tilde s^z_{k+r} | \Psi \rangle \nonumber ~~.
\end{eqnarray}
We see from Eq. (\ref{appendix_2}) above that this expression
may be written in terms of our high-order CCM operators as
\begin{eqnarray}
M_{r}^{zz} &=& \langle \Phi | \bigl \{ \frac 1N + \sum_{I=1}^{N_F} 
\frac {\tilde x_I}{A_I}  C_I^- \bigr \} \sum_{k=1}^N
\biggl ( s^z_k  s^z_{k+r} + F_m s_{k+r}^+ s_k^z 
+ F_k s_k^+ s_{k+r}^z  \nonumber \\
&&~~~~~~~~~~ + G_{k,k+r} s_k^+ s_{k+r}^+ + F_k F_{k+r}  
s_k^+ s_{k+r}^+ \biggr ) 
| \Phi \rangle ~~ .
\label{appendix_3}
\end{eqnarray}
The right-hand-side of Eq. (\ref{appendix_3}) may be evaluated 
computationally in exactly the same manner as 
for the ket-state equations, 
although no results for spin-spin correlation functions are quoted 
in this article.

\pagebreak


\pagebreak
\begin{figure}
\epsfxsize=13cm
\centerline{\epsffile{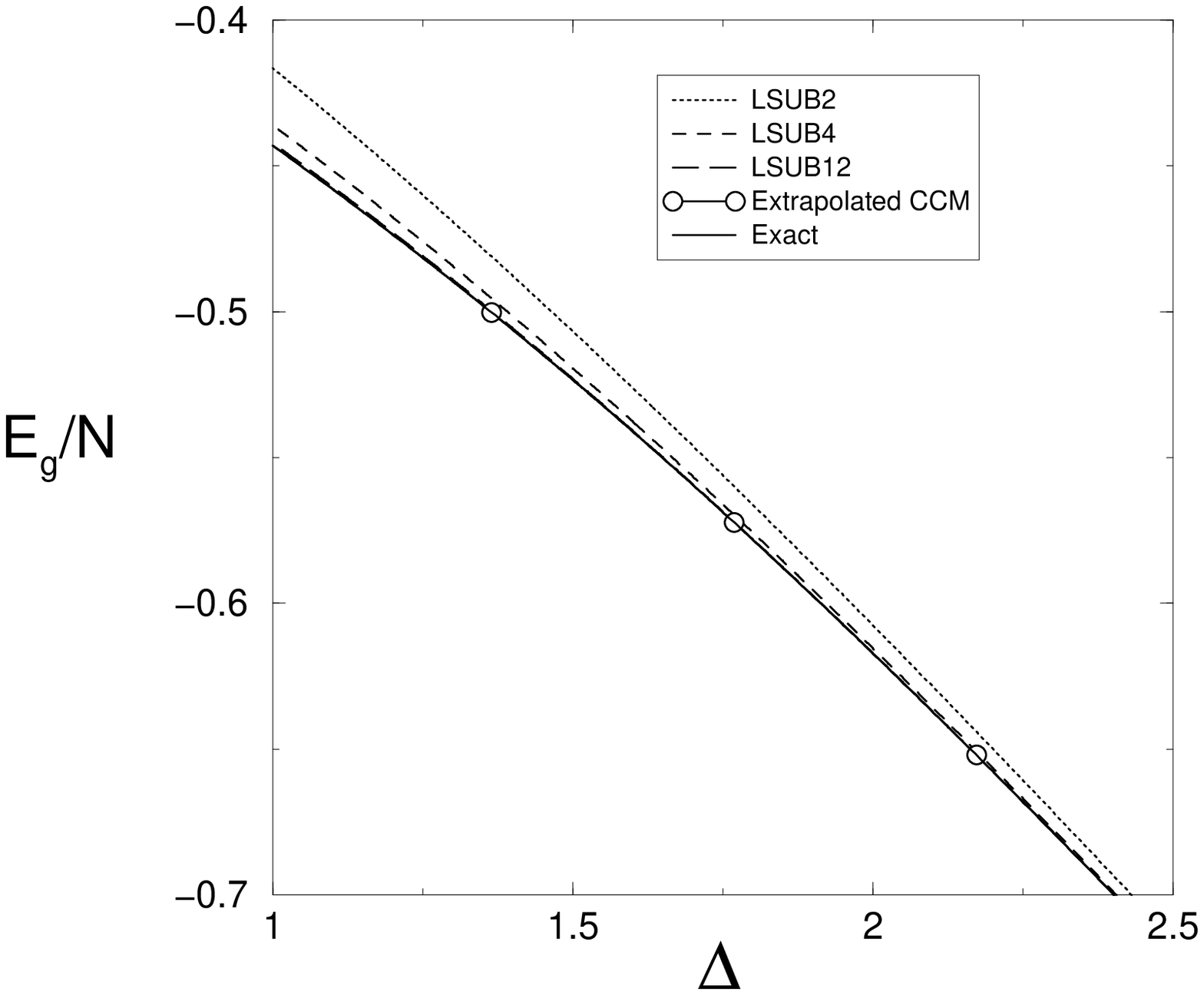}}
\vspace{0.4cm}
\caption{CCM results for the ground-state energy per spin of the 
spin-half {\it XXZ} model on the linear chain using the LSUB$m$ 
approximation scheme with $m=\{2,4,12\}$. The LSUB$m$ results for 
$m=\{6,8,10,12\}$ are extrapolated in the limit $m \rightarrow 
\infty$ for this case and are compared to exact results of the 
Bethe Ansatz\cite{ba1,ba2,ba3,ba4}.  (Note that the LSUB$m$ and 
SUB$m$-$m$ approximation schemes are equivalent for this model.)}
\label{fig1}
\end{figure}

\pagebreak
\begin{figure}
\epsfxsize=13cm
\centerline{\epsffile{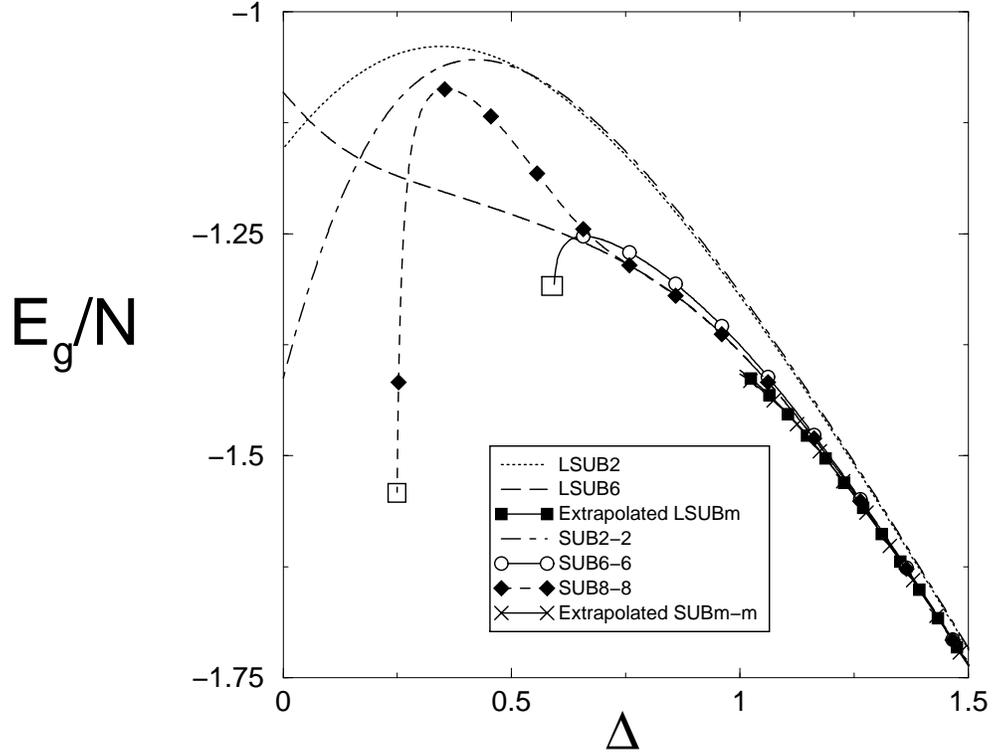}}
\vspace{0.4cm}
\caption{CCM results for the ground-state energy per spin of the 
spin-one {\it XXZ} model on the linear chain using the SUB$m$-$m$ 
approximation scheme with $m=\{2,6,8\}$ and the LSUB$m$ 
approximation scheme with $m=\{2,6\}$. LSUB$m$ and SUB$m$-$m$ 
results are extrapolated  in the limit $m \rightarrow \infty$ 
for $\Delta \ge 1$. Note that boxes indicate the CCM critical points.}
\label{fig_energy_spin_one}
\end{figure}

\pagebreak
\begin{figure}
\epsfxsize=13cm
\centerline{\epsffile{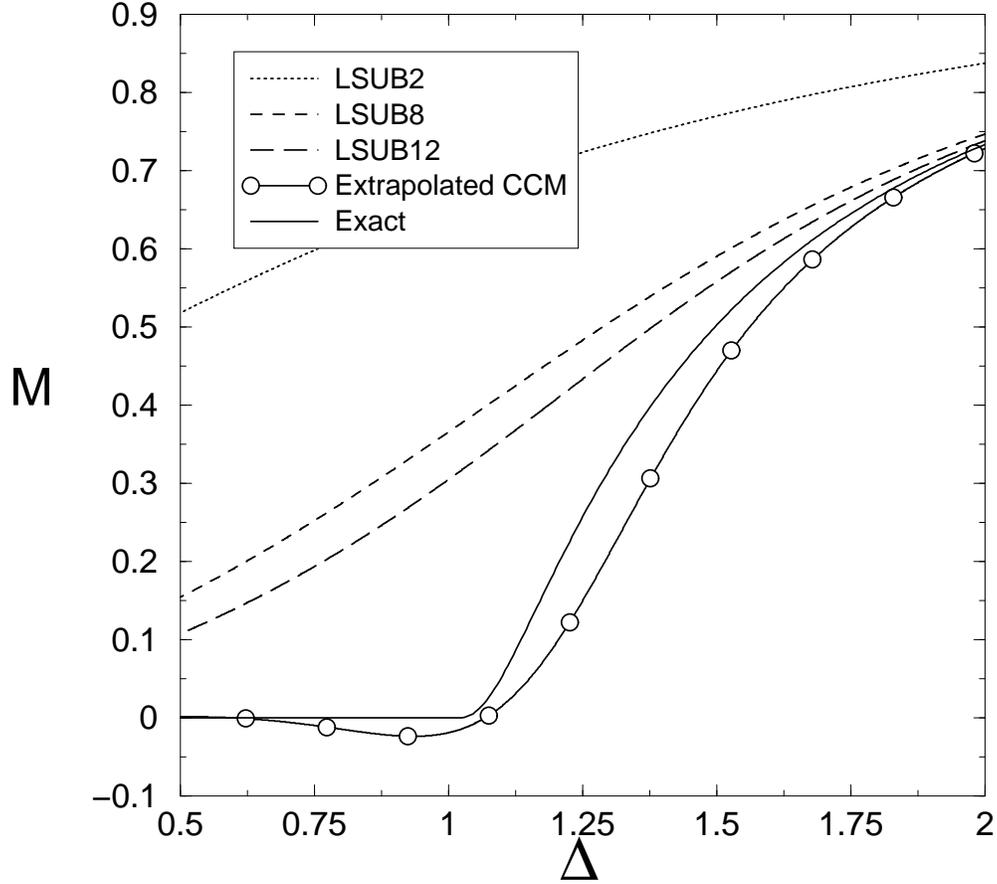}}
\vspace{0.4cm}
\caption{CCM results for the sublattice magnetisation of the 
spin-half {\it XXZ} model on the linear chain using the LSUB$m$ 
approximation scheme with $m=\{2,8,12\}$. The LSUB$m$ results for 
$m=\{6,8,10,12\}$ are extrapolated in the limit $m \rightarrow 
\infty$ and are compared to exact results of Bethe 
Ansatz\cite{ba1,ba2,ba3,ba4}. Note that the 
extrapolated CCM results follow the qualitative behaviour
of the exact results quite closely and that the 
extrapolated results go to zero at $\Delta=1.07$.}
\label{fig2}
\end{figure}

\pagebreak
\begin{figure}
\epsfxsize=13cm
\centerline{\epsffile{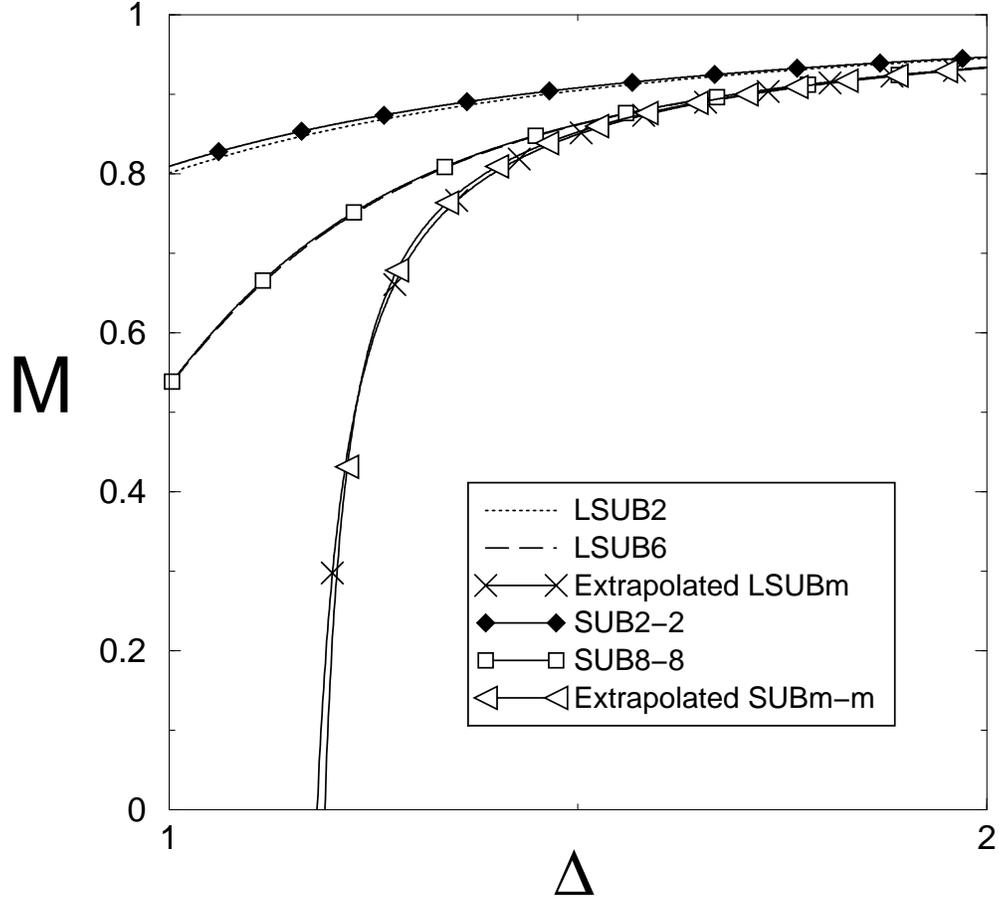}}
\vspace{0.4cm}
\caption{CCM results for the sublattice magnetisation of the 
spin-one {\it XXZ} model on the linear chain using the LSUB$m$ 
approximation scheme with $m=\{2,6\}$ and SUB$m$-$m$ approximation scheme with 
$m=\{2,8\}$. Results for the LSUB$m$ approximation scheme with $m=\{2,4,6\}$ 
and SUB$m$-$m$ approximation scheme with $m=\{2,8\}$ are extrapolated 
in the limit $m \rightarrow \infty$. Note that the 
extrapolated CCM results go to zero at $\Delta \approx 1.19$ and 
$\Delta \approx 1.18$ for the SUB$m$-$m$ and LSUB$m$ approximation schemes,
respectively.} 
\label{fig3}
\end{figure}

\pagebreak
\begin{figure}
\epsfxsize=13cm
\centerline{\epsffile{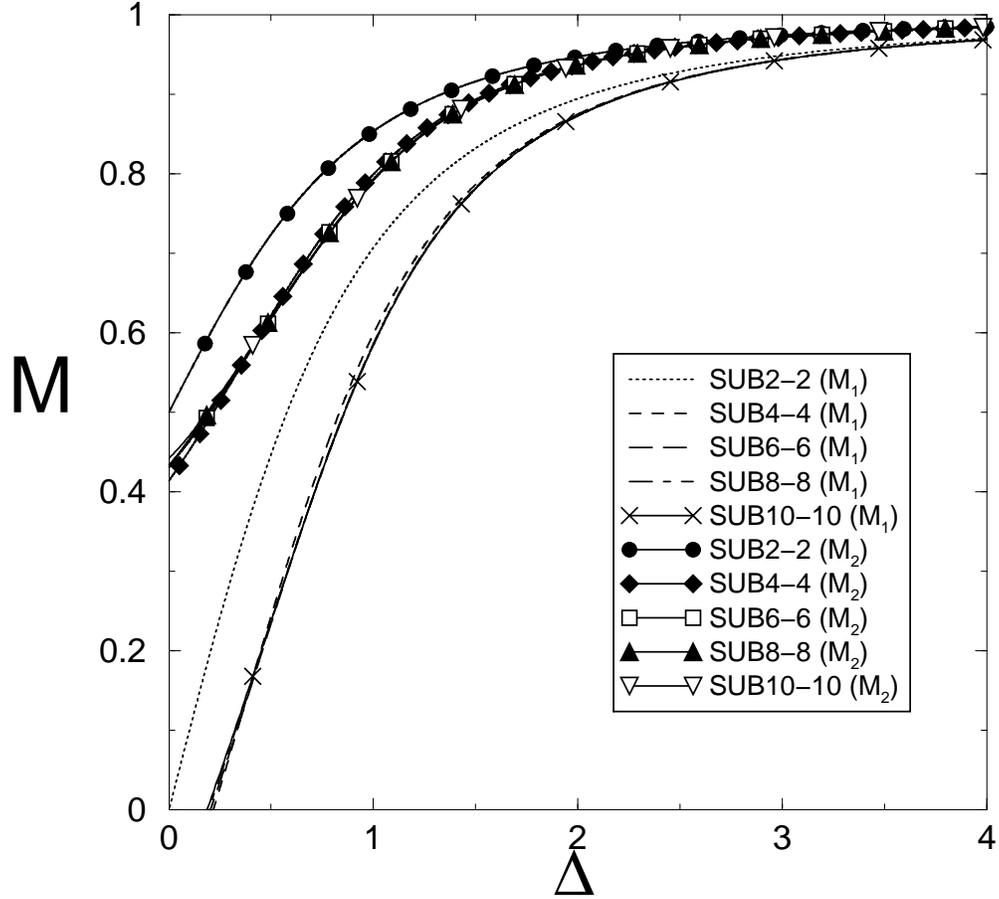}}
\vspace{0.4cm}
\caption{CCM results for the sublattice magnetisation 
of the spin-half spins, $M_1$, and of the spin-one spins, $M_2$, 
for the spin-half/spin-one {\it XXZ} ferrimagnet on the linear chain 
using SUB$m$-$m$ approximation scheme with $m=\{2,4,6,8,10\}$. Note that 
the LSUB$m$ and SUB$m$-$m$ approximation schemes are equivalent 
for this model at the levels of truncation index $m$ shown.}
\label{fig4}
\end{figure}

\pagebreak


\begin{table}
\caption{Results obtained for the spin-half Heisenberg model on the 
linear chain using the CCM LSUB$m$  approximation 
scheme with $m=\{2,4,6,8,10,12\}$. 
$N_{F}$ denotes the number of fundamental configurations for the 
ground state. The ground-state energy per spin, $E_g/N$, and the 
sublattice magnetisation, $M$ are shown. The LSUB$m$ results for 
$m=\{6,8,10,12\}$ are extrapolated in the limit $m \rightarrow 
\infty$ for this case and are compared to exact results of the 
Bethe Ansatz\cite{ba1,ba2,ba3,ba4}. (Numbers in brackets for
the extrapolated CCM results indicate the estimated error in
the last significant figure shown. Note that the LSUB$m$ and 
SUB$m$-$m$ approximation schemes are equivalent for this model.)}
\begin{center}
\begin{tabular}{|l|c|c|c|}  \hline\hline
        &$N_{F}$    	 &$E_g/N$ 
        &$M$            \\ \hline\hline
LSUB2   &1              &$-$0.416667 
        &0.666667       \\ \hline
LSUB4   &3              &$-$0.436270
        &0.496776       \\ \hline
LSUB6   &9              &$-$0.440024
        &0.415771       \\ \hline
LSUB8   &26             &$-$0.441366
        &0.365943       \\ \hline
LSUB10  &81             &$-$0.441995  
        &0.331249       \\ \hline
LSUB12  &267            &$-$0.442340
        &0.305254       \\ \hline
Extrapolated CCM 
        &--             &$-$0.44315084(6)
	&$-$0.01876(3)  \\ \hline\hline
Bethe Ansatz&--    	&$-$0.443147    
        &0.0            \\ \hline
\end{tabular}
\end{center}
\vspace{20pt}
\label{tab1}
\end{table}

\pagebreak

\begin{table}
\caption{Results obtained for the spin-one Heisenberg model on the 
linear chain using the LSUB$m$ approximation scheme with $m=\{2,4,6\}$ and 
SUB$m$-$m$ approximation 
scheme with $m=\{2,4,6,8,10\}$. $N_{F}$ denotes the number 
of fundamental configurations for the ground state. The ground-state 
energy per spin, $E_g/N$, the sublattice magnetisation, $M$, and the critical 
values of $\Delta_c$ are shown. The SUB$m$-$m$ results for 
$m=\{4,6,8\}$ and LSUB$m$ results for $m=\{2,4,6\}$ are 
extrapolated in the limit $m \rightarrow \infty$ in this case, 
and are compared to results$^{\dag\dag}$ of the DMRG method \cite{DMRG4}. 
The ** symbol indicates that extrapolated CCM results for the sublattice 
magnetisation, $M$, go to zero at $\Delta_c = 1.19$ and
$\Delta_c = 1.18$ for the SUB$m$-$m$ and LSUB$m$ approximations, 
respectively, ($^{\dag}$ also given in this table) and are thus 
ill-defined below these points. The result$^{*}$ for the position
of the phase transition point using the large-cluster-decomposition
Monte Carlo method \cite{Nomura} is also quoted.}
\begin{center}
\begin{tabular}{|l|c|c|c|c|}  \hline\hline
        &$N_{F}$         &$E_g/N$       &$M$       &$\Delta_c$  
\\ \hline\hline
SUB2-2   &1               &$-$1.316625   &0.809068  &--    \\ \hline
SUB4-4   &7               &$-$1.360084   &0.694610  &--    \\ \hline
SUB6-6   &37              &$-$1.375607   &0.607339  &0.593 \\ \hline
SUB8-8   &247             &$-$1.383466   &0.533252  &0.249 \\ \hline
Extrapolated SUB$m$-$m$ 
         &--              &$-$1.408039   &**        &1.19$^{\dag}$\\ \hline
\hline
LSUB2    &2               &$-$1.320608   &0.800702  &--\\ \hline
LSUB4    &11              &$-$1.369428   &0.646536  &--\\ \hline
LSUB6    &63              &$-$1.383292   &0.532198  &$-$0.670\\ \hline
Extrapolated LSUB$m$  
         &--              &$-$1.403737   &**        &1.18$^{\dag}$\\ \hline
\hline
c.f.     &--              &$-$1.401484038971(4)$^{\dag\dag}$   
         &0.0             &1.167(7)$^{*}$ \\ \hline
\end{tabular}
\end{center}
\vspace{20pt}
\label{tab2}
\end{table}

\pagebreak

\begin{table}
\caption{Results obtained for the spin-half/spin-one ferrimagnetic HAF on the 
linear chain using the CCM SUB$m$-$m$ approximation scheme with $m=\{2,4,6,8,10\}$ 
compared to results of DMRG \cite{DMRG5} calculations.  
CCM results for $m=\{6,8,10\}$ are extrapolated in the limit 
$m \rightarrow \infty$. $N_{F}$ denotes the 
number of fundamental configurations for the ground state. The ground-state 
energy per spin, $E_g/N$, the sublattice magnetisation on the spin-half 
sites, $M_1$, the sublattice magnetisation on the spin-one sites, $M_2$
are shown. (Note that the LSUB$m$ and SUB$m$-$m$ approximation schemes are 
equivalent at the levels of approximation shown for this model.)}
\begin{center}
\begin{tabular}{|l|c|c|c|c|}  \hline\hline
          &$N_{F}$ &$E_g/N$         &$M_1$     &$M_2$      \\ 
\hline\hline
SUB2-2    &1       &$-$0.70710678   &0.70710678  &0.85355339 \\ \hline
SUB4-4    &5       &$-$0.72582592   &0.59865621  &0.79932811 \\ \hline
SUB6-6    &21      &$-$0.72697237   &0.58611255  &0.79305628 \\ \hline
SUB8-8    &93      &$-$0.72704344   &0.58503667  &0.79251834 \\ \hline
SUB10-10  &427     &$-$0.72704696   &0.58497261  &0.79248630 \\ \hline
Extrapolated CCM  
          &--      &$-$0.7270474    &0.5849641   &0.7924820  \\ \hline\hline 
DMRG      &--      &$-$0.72704      &0.58496     &0.79248    \\ \hline
\end{tabular}
\end{center}
\vspace{20pt}
\label{tab3}
\end{table}

\end{document}